# Analysing GSM Insecurity

Ene Donald, Osagie Nosa Favour

*Rivers State University of Science and Technology, Nigeria*

*Abstract:* In the 1990s, GSM emerged as a cutting-edge technology that promised improved services, mobility, security, and increase in the revenues of companies through improved, secure communication for business transactions. The practical experience, however, has shown global subscribers and companies that mere technological excellence of GSM has not resulted in its success. This paper has been carried out to analyse the existing GSM subscriber activities in both personal and business worlds by implementing a customized GSM baseband protocol stack on a phone. The paper encompasses the tasks of analysing the prevailing GSM insecurity, establishing a conceptual design towards the findings revealed in the practical personal and business environments and knowledge gathered from the literature.

*Key Words:* GSM, Technology, GSM Activities, GSM Insecurities.

## I. INTRODUCTION

Mobile phones have become components of our daily lives that we very much depend on; in the near future, we will depend on them more than computers. Mobile phones use GSM technology, which is an open digital cellular technology that transmits voice, mobile, and data service on the GSM network. This paper is about the analysis of GSM insecurity. Both GSM and TCP/IP are publicly available; there is no difference on the availability of public information on both protocols. The Internet protocol stack (Ethernet/WiFi/TCP/IP) receives lots of scrutiny from the independent hackers community, academic community, and everyone has looked into that protocol security issue for more than 10 years. The GSM network is as widely deployed as the Internet with a user base of more than 3 billion and still counting (Mulliner & Golde, 2010), yet the GSM protocol stack has not received such scrutiny and research. This is because the GSM industries do not allow the availability of information. They are rather extremely closed. According to Steve Margrave, (n.d), there are only a few (say about 4) closed-source protocol stack implementations available from IP2 lower number. Secondly, the chipset manufacturers that produce the baseband chipsets do not release any hardware documentation publicly. They only release it to the actual cell phone manufacturers in a limited fashion. The companies that manufacture the baseband chipsets are;

1. Texas Instruments
2. Infineon
3. Mediatek
4. Qualcomm
5. ST-Ericson
6. Freescale
7. Broadcom

These companies often license GSM protocol stack and operating system kernel from third parties (Weinmann, 2009). Only a few handset makers are large enough to become customers. These customers only have limited access to hardware documentation and no access to firmware source.

## II. LITERATURE REVIEW

As earlier said, mobile phones have become components of our daily lives that we very much depend on; we will depend on them more than computers in the near future. Hand phones use GSM technology to transmit voice and data services over the GSM network. However, GSM network pose serious threat to the users. These threat are born out of the security consequences and negligence on the side of the closed GSM industry.

Due to the extremely closed state of the GSM industry for the past decade, there has not been any independent practical research on the GSM protocol-level security. Therefore, apart from a couple of projects like the TSM30 – which was part of the GSM project – and also Mados – an alternative operating system for Nokia DTC3 phones - that have attempted to solve this problem but without success, there was no open source protocol implementation. Open Base Transceiver Station (OpenBTS), Open Base Station Controller (OpenBSC), and Open Source Mobile Communication Baseband (OSMOCOMBB) are the only open source protocol implementation projects that have tried and succeeded on practically researching on the GSM protocol-level implementation. Hence, there is a limited number of documents to review on this topic. On the security aspect of the GSM network, however, there are a couple of them.

A search for GSM Security return over 4.8 million references, but due to acute time constrain, 6 of those references from different authors were read for this paper. A close look revealed that the different authors are talking about pretty the same thing, except for a few differences projected by some authors. Of all the references studied for this paper so far, the following was obtained in abstract form from some authors. Gadaix, E. (2001) explained the following to be the problems with GSM Security;

- They only provide access security – communications and signalling traffic on the fixed networks are not protected
- The devices are only as secured as the fixed networks to which they connect
- Active attacks on the networks are not addressed
- Lawful interception are only considered as an after-thought

- It is difficult and near impossible to upgrade the cryptographic mechanisms
- Terminal identity cannot be trusted
- Lack of user visibility

*Findings*

*Architecture Overview*

The GSM network is made up of a few functional entities with functions and well-defined interfaces. The GSM system is an open, digital cellular technology used to transmit voice and data services (TutorialsPoint, 2011). GSM uses narrowband TDMA technique to transmit signals. According to TutorialsPoint, GSM provides basic to advanced voice and data services including roaming services. Operating at either 900MHz or 1800MHz frequency band, the GSM system digitizes and compresses data. Then the compressed data is brought down through a channel with two other streams of user data each in its own time slot.

TutrialsPoint explained that the GSM was introduced to solve a couple of problems. These made some enhancements in the phone system. The enhancements include;

- International roaming
- Improved spectrum efficiency
- Compatibility with ISDN and other telco services
- High quality speech
- Low cost mobile equipment and base station
- Support for new services

The methods of security standardized for GSM system at the time (in the 90s) made it the most secured cellular telecommunications standard available then, but currently, it is probably obsolete. Since the confidentiality and a call and the anonymity of a subscriber was promised on the radio channel (Um interface), it becomes a major problem in achieving end-to-end security (Margrave, n.d).

According to Margrave, n.d, the additional parts or elements of the GSM architecture comprises of the function of the database and messaging systems:

- Authentication Center (AuC)
- Home Location Register (HLR)
- Equipment Identity Register (EIR)
- Chargeback Center (CBC)
- SMS Serving Center (SMS SC)
- Visitor Location Register (VLR)
- Gateway MSC (GMSC)
- Transcoder and Adaptation Unit (TRAU)

Steve also explained that the security consequences of the closed GSM industry include the fact that

I. There are no independent protocol level research except for the theoretical research like the A5/1 and A5/2 cryptanalysis.
II. Other than the protocol or GSM network equipment manufacturers, no one has any detailed technical knowledge of what is actually going on.
III. Before the implementation of the open base transceiver station (OpenBSC), open base station controller (OpenBSC), open source mobile communication baseband (OSMOCOMBB) approximately three years ago, there was no open source protocol implementation.

These projects, OsmocomBB, OpenBSC, and OpenBTS have helped people learn about the protocols. The big question is how does one start with the GSM protocol level security analysis?

*The Network Side*

It is pretty difficult to start the security analysis on the network side, since the network equipments are not easily available and are usually highly expensive. If the GSM network equipments were available and affordable, security analysis would be enhanced. On the other hand, networks are very modular and have many standardized documented interfaces. OpenBTS and OpenBSC projects have already done the analysis on the network side in 2008/2009 (Margrave, n.d).

*The Handset Side*

The handset side has its own difficulties as well, some of which include the fact that GSM firmware are proprietary and closed. On the thought of anyone writing a custom protocol stack, the layer1 hardware and signal processing is undocumented and closed as well. A couple of projects like the TSM30 which was part of the GSM project and also Mados, an alternative OS for Nokia DTC3 phones, have tried solving this problem but didn't come out successful (Margrave, n.d).

For the purpose of this security analysis the GSM system would be considered in parts.

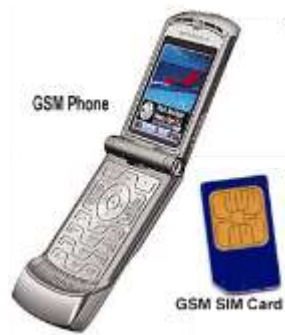

Figure 2.1: The Mobile Station
Image Source [15]

*The Mobile Station*

*Definition*

The mobile station (ME) is a combination of physical terminal equipment such as the radio transceiver, the digital signal processor, the display, and the SIM card where the subscriber data is stored. Therefore, (ME + SIM = MS)(Nokia, 2002).

The MS enables the user to connect to the air interface (Um) of the GSM network. In addition to this, other services provided include data bearer services, voice teleservices, and feature supplementary services. Among other functions, the mobile station has an SMS receptor, enabling user to toggle between the voice and data use. The MS helps access to voice and messaging systems. It also facilitates the access to a high speed, circuit switch data at speeds up to 64Kbps, X.25 switching through an asynchronous and synchronous dialup connection to pad at 9.6Kbps, GPRS using either X.25 or IP based data transfer method (TutorialsPoint, 2011).

The SIM contains user-specific identification. As a small memory mounted on a card, the SIM can be removed from one ME to another without losing its data. For security reasons, the SIM has the tools needed for authentication and cyphering (Nokia, 2002). The SIM provides a medium for the user to access all subscribed services not taking into account the location of the terminal and the use of a specific terminal (TutorialsPoint, 2011).

*The Base Station Subsystem (BSS) Architecture*

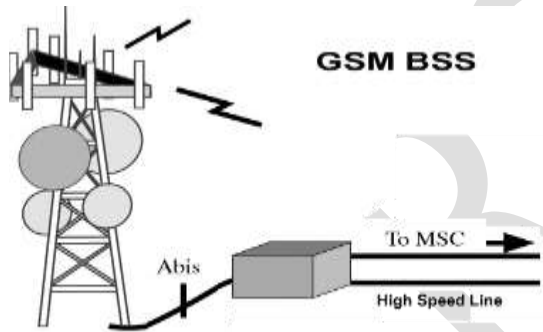

Figure 2.2: The BSS
Image Source [15]

Following the definition given by Markgrave, "the base station subsystem (BSS) is a set of equipment (Aerials, transceivers, and a controller) that is viewed by the mobile switching centre (MSC) through a single A-interface as being the entity responsible for connecting with mobile station in a certain area." He also said that the BSS is made of two parts;

- The Base Transceiver Station (BTS)
- The Base Station Controller (BSC)

The BSC and the BTS communicate across a specified A-bis interface enabling components that are made by different suppliers to operate together. The A-bis interface is the interface that lies between the BSC and the BTS. The radio interface, which is the main focus of this project, is where the contact is made between the MS and the BTS. The radio interface is sometimes encrypted and most times when this encryption fails, user data and voice through the radio interface become unencrypted (Steve, 2009).

*The Base Transceiver Station (BTS) Architecture*

The BTS is the cell tower that is responsible for the transmission and reception of the radio signal from the MS over the air interface. It marks the boundary and limit of a cell and controls the radio link protocol with the MS. Placed in the centre of a cell, the transmitting power of the BTS sets the limit of a cell. According to the account of Steve, 2009, Depending on the user density of a cell, each BTS has between 1 and 6 transceivers. The BTS is responsible for the following:

- Transcoding and rate adaptation
- Random access detection
- Voice through full or half-rate service
- Timing advances
- Uplink channel measurement
- Time and frequency measurement
- Encrypting, decoding, multiplexing, modulating and sending the radio frequency signal to the antenna (TutorialsPoint, 2011).

*The Base Station Controller (BSC) Architecture*

The BSC handles the radio resource (RR) management for one or more BTSs. It allocates and releases frequencies for all MSs in a given area (Steve, 2009). The BSC is responsible for channel setup, frequency hopping, and inter-cell handovers. The BSC is responsible for the translation of 13Kbps voice channel used over the radio link to the standard 64Kbps used in PSTN.

*The Network Switching Subsystem (NSS) Architecture*

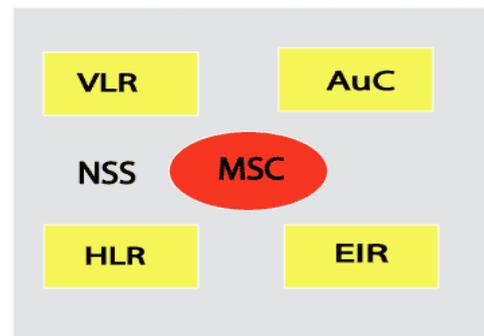

Figure 2.3: The BSS
Image Source [15]

According to TutorialsPoint, the NSS has 5 components;

*MSC:* The core part of the NSS is the mobile switching centre. The job of the MSC is to switch calls between mobiles or other mobile networks and fixed line users, as well as managing the mobile services such as authentication.

*HLR:* The home location register is the database used to store and manage subscriptions. As the most important database, it stores permanent data about subscribers and their service profile, location information, and their activity status.

*VLR:* The visitor location register is another database, but this time, it temporarily stores information of subscribers that is needed by the MSC so as to service visiting subscribers. When a new mobile station roams into a new MSC area, the VLR will request data about the mobile station from the HLR. If the mobile station makes a call, the VLR will have the information needed for call setup instead of querying the HLR every time.

*AUC:* The authentication centre is the protected database that stores a copy of the secret keys that are stored on the SIM card. These secret keys are used to authenticate and cipher the radio channel.

*EIR:* The equipment identity register is another database that contains the list of all valid mobile equipment on the network, where the mobile stations are identified by its international mobile equipment identity (IMEI).

### III. CASE STUDY

The literature presents the description and definition of some security threats on the GSM network and the possible risks involved resulting from mobile phone subscribers not being allowed to manage their networks like the Internet users.

The key concern of this paper is that security is a multi-layered problem and the strength of the solution is dependent on the security of the weakest link (Um interface). The derivation of the requirement for this complex problem was from the security concern and requirement of the subscribers across different networks. Their responses were analysed through use and misuse cases spanning all the interfaces of the multi-layered architecture to determine attacks and security requirements.

Finally, this project attempts to give hackers and the independent researchers a clue of the GSM protocol stack, a fair chance of discovering bugs, scrutinizing the protocol stack where the response by the GSM industry would create a safer GSM network communication, and counting the security implication resulting from the closed-mindedness of the GSM industry.

### A. Background

The GSM network is made of cells that receive signals from and transmit signals to mobile stations. Between the mobile station and the BTS is the air (Um) interface. As explained above, the GSM protocol stack is always connected to the GSM network through the baseband processor (BP). The BP runs the digital signal processing for RF layer 1. As long as the mobile station is always connected to the GSM network, subscribers face serious security threats. According Walte (2009), the things that give rise to the threats include;

(i) As weak as the encryption algorithm is, it is optional, users never have an idea if it is active or not.
(ii) There is no mutual authentication between the network and the mobile phone which could give lead to a number of attacks, like rogue network attacks, man-in-the-middle attacks, it also what enables IMSI-catchers.
(iii) DoS attack of the RACH by means of channel flooding.
(iv) With the Radio Resource Locator Protocol (RRLP), the network can obtain a GSM fix or even a raw GSM data from the phone, combine with the fact the network does not need to authenticate itself.
(v) The software stack on the baseband processor is written in C and assembly languages. As we all know, these low level languages lack any modern security features like stack protection, non-executable pages, address space randomization, etc.

However, as earlier mentioned, the goal of this paper involves giving hackers and the independent researchers a clue of the GSM protocol stack, a fair chance of discovering bugs mentioned above, a chance to scrutinize the protocol stack where the response by the GSM industry would create a safer and more secure GSM network communication rather than assuming that obscurity saves the matter.

### B. Analysis

This section introduces the basic methodology that was employed for the elicitation of the requirement from the mobile subscribers. It also introduces the overall approach that was employed to establish the analysis, requirements and recommendations.

*1. Approach*

*Overall Process Model*

One of the goals of this paper includes assessing and anticipating the security needs of various subscribers as an input for enabling an open and secure GSM network. Fig. 1 illustrates the overall process in the direction of security requirement and recommendation. This figure is made up of a set of activities and result that are connected to each other. The description of the activities such as scenarios, interviews, and requirement elicitation are presented subsections of this paper.

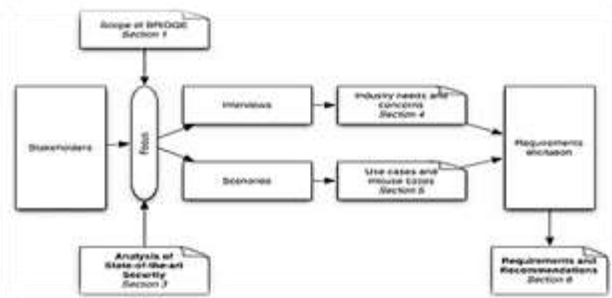

Figure 3.1. Overall Approach of requirement gathering
Image Source [18]

This paper will analyse the security implications of the closed GSM industry and how it affects the hackers' community, the independent researchers, and the subscribers. As the target of this project involves making the baseband protocol stack an open source technology, a two-phase approach will be used to gather information on their security needs and concerns. The first phase would involve conducting interviews with organizations such as the banks and the transport companies that use GSM technology, and provision of questionnaires for individual subscribers as well. This large subscriber base accounts for almost 80% of GSM usage. The second phase involves creating use case scenarios in which an attacker can jeopardize the system. Based on these two kinds of inputs, their security needs and concerns would be established deriving the requirements according to the selected proven methodology.

*2. Data Gathering Approach*

In fulfilment of the goal of assessing the requirement of the vast number of subscribers and multiple industrial users, the choice of data gathering hovered between online survey and interview. After the reconsideration of the goal of this paper, the idea of the web-based was discarded and replaced with precise questionnaire. The replacement was due to the fact that this paper targets a potential solution in the future rather the experience with the existing flaws. Therefore, the data gathering strategy focuses on the qualitative exploration of knowledge instead of the quantitative approach. Hence, an organized approach of interviews with experts, and industrial users, as well as precise questions on the questionnaire for individual subscribers was deemed appropriate. Since the GSM technology is the same everywhere, the interviews were carried out in two countries (Singapore and Nigeria) representing a first world and developing countries respectively. To assist the explorative quality of project, the modes of interview were partly by phone and emails in Singapore and chiefly by email correspondence in Nigerian due to the distance. The mode of data gathering with the questionnaire was on one-on-one bases from individual subscribers.

*3. Scenarios and Requirement Methodology*

Besides conducting the interviews and questionnaires, another approach considering the use case scenarios will be adopted, in which the GSM network infrastructures will be utilized and the possible ways in which malicious user or attacker can endanger the system. There are two-fold reasons for using this approach. The first would be to present a scenario from the solution that is intended to be achieved by this project, trying to intentionally imagine what threats could be faced. The second fold is from a set of reference scenarios from the GSM communication and network services from different service providers on a common vocabulary. Common use and misuse cases will be used to derive different security requirements according to their interests. The remaining part of this section will describe the methodology used to obtain security requirements from use case scenarios.

A narrative of each scenario will be given, with the required background and probably any possible assumptions. The description may be divided into different scenes, which are the building blocks for the scenarios. From the description, a use case diagram will be derived that summarizes the narrative in one picture. Then the requirements elicitation begins with the definition of possible misuse cases that the attacker or malicious user can carryout to endanger the system. These misuse cases require certain security requirements to mitigate the threats described in the scenario. These requirements can be directly derived typically from securing the use cases. Then the result will picture the complete use and misuse cases.

*C. The GSM Architecture and Current Security Capabilities*

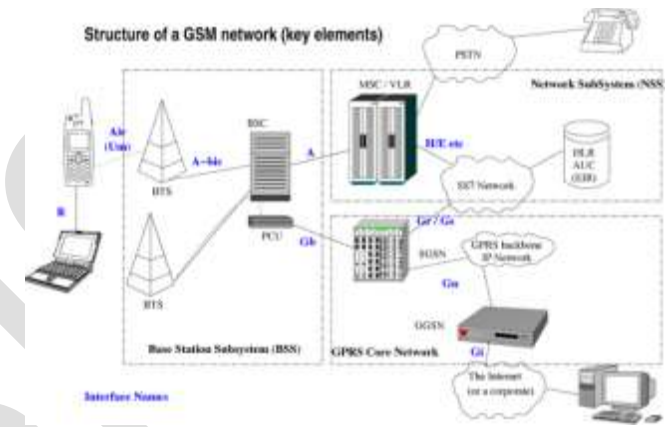

Figure 3.2: GSM Network Interface Architecture
Image Source [19]

*1. Security on the public network*

Just as well GSM, TCP/IP are publicly available; there is no difference on the availability of public information on both protocols. The Internet protocol stack (Ethernet/TCP/IP/WIFI) receives lots of scrutiny from the independent hackers' community, academic community, and everyone has looked into the protocol security issue for the past decade. The scrutiny of the Internet protocol stack has been yielding some very positive results in terms of fixing the bugs that are security relevant and updates.

Unlike the security of the GSM network, the security in the Internet is managed and ensured by the user with the help of the computer security industries. According to Microsoft Safety and Security Centre (2011), there are four quick tips to enjoy the convenience of public WIFI and help to protect your privacy;

- Use a firewall
- Hide your files
- Do not type in credit cards numbers or passwords
- Turn off your wireless network when you are not using it

Although this is not the area of interest to be considered in this paper, but the review looks to analyse and compare the GSM protocol stack to the Internet protocol stack. To maintain confidentiality and availability on the public network, Rose India (2008) suggested some precautionary measures of protection while using a public network;

- Use legitimate access points
- Encrypt your sensitive files
- Use VPN
- Setup a personal firewall
- Install and regularly update your security software
- Update your operating system
- Ensure privacy
- Disable file sharing
- Protect your confidential files with strong passwords

*2. GSM Network Security*

According Walte, H. and Markgraf, S. (2009), "Every mobile device that is connected to a cellular network runs some kind of baseband processor with highly proprietary and closed-source firmware." These authors also explained that every reasonably complex, security relevant software has bugs that might be exploited. It has been a notable fact that "there is no security in obscurity", open source projects provide a higher level security compared to closed source projects. This is because as more eyes review the open source codes, security related bugs are fixed almost immediately.

Everyone that use computer understands that connecting a computer that is not protected to a public network like the Internet is dangerous. As a result of this, people use their application level gateways, dedicated or personal firewalls, virus scanners, and other technologies to protect themselves. But the mobile phone is different, specifically the baseband processor that is permanently attached to a public network. In direct contrast to computer and public network where the security of the user information is managed and ensured by the user. Karsten Nohl & Sylvain Manaut, (2010) says that the GSM network is

- o Managed and controlled by the service providers and hardware manufacturers
- o No incident response management
- o No clean way bugs can be fixed and updated quickly
- o The device manufacturers rarely release firmware updates
- o No software protection; the entire software stack is run in supervisor mode
- o GSM operators do not authenticate each other but leak subscribers' private data

Due to these problems with GSM security, the following are the possible threats or attacks.

*Eavesdropping:* An attacker can eavesdrop on signals and data connection associated with another user or target because of weak encryption and the intermittent security failure. This can be done with a modified mobile station (MS) Gadaix, E. (2001). It happens as a result of weak or no authentication of messages received over the radio interface. The target will be enticed to connect to on a false network set up with a fake Base Transceiver Station.

*Impersonation of the network:* The attacker can send signals and or user data to a target making the target believe they are from a genuine network. This is possible because there is no mutual authentication between the mobile station and the GSM network. This attack can be carried out using a modified base transceiver station (BTS) Gadaix, E. (2001).

*Impersonation of a user:* In this attack, the attacker can send signal and or user data to the network, tricking the network to believe that the signal originated from the target user. This is due to the weakness in encryption that makes SIM card cloning possible. This attack can be carried out with a modified mobile station Gadaix, E. (2001). This happens when there is no mutual authentication between the network and the mobile station.

*Man-in-the-middle:* On this, the attacker positions itself between the genuine network and the targeted user in an attempt to reply, re-order, modify, delete, and spoof signal, messages and user data exchanged between the network and the target user. This attack can be carried out with a modified base transceiver station and mobile station Gadaix, E. (2001). This can happens when the phone and network do not authenticate each other.

*Compromising authentication vectors on the network:* An attacker can possess an authentication vector that has been compromised. This might include challenge/response pairs, integrity keys, and cipher keys. These can be obtained by compromising network nodes or by obtaining signalling messages on network links Gadaix, E. (2001).

*3. Security Capabilities*

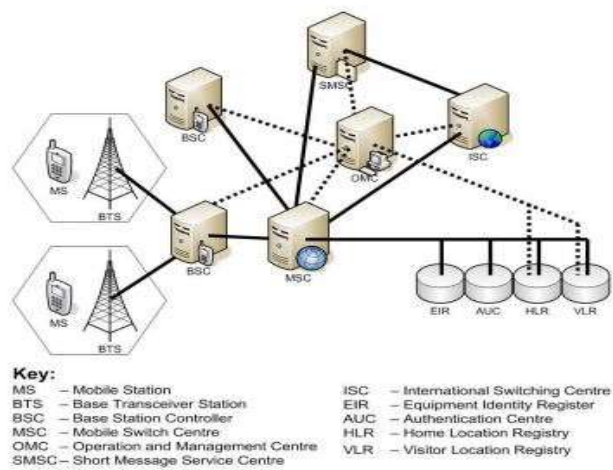

Figure 3.3: GSM Network Security Architecture
Image Source [5]

*4. GSM Security Mechanism*

The details of the GSM security aspects are given in the GSM Recommendations 02.09 (Security Aspects), 02.16 (International Mobile Equipment Identity), 02.17 (Subscriber Identity Module), 03.20 (Security-Related Network Functions). As stated above, security mechanism in the GSM system is a multi-layer process implemented in different system elements, therefore, GSM security is made of the following layers; subscriber identity confidentiality, subscriber identity authentication, user data confidentiality, and signaling data confidentiality (Markgrave, 2009). A unique International Mobile Subscriber Identity (IMSI) identifies every subscriber. Along with the individual subscriber authentication key (Ki), this information constitutes sensitive identification credential comparable in certain respect to the Electronic Serial Number (ESN) in the analog systems. The real conversations are encrypted using a temporary randomly generated cypher key (Kc) that always fails.

To provide mobile communication privacy, an A5 stream cipher is provided on every MS. It is a combination of either four (A5/2) linear feedback shift registers (LFSRs) by which data are encrypted or decrypted or three (A5/1) ciphers. The A8 (operation proprietary algorithm) on the SIM generates a random input known as RAND that creates the session key (Kc) and uses this algorithm. The A8 session keys that are used to encrypt SMS and calls are used once in every session, after which they are recycled a couple of times and a new key is generated. The session key can be predicted by XOR-ing the known key to get the next one. (Sylvain, 2010). The RAND generated by the SIM is the same used in the authentication process. The MS specifies the supported encryption algorithm. Then with the cipher mode command, the BTS informs the MS the chosen one to use (Stobel, 2007).

However, the authentication process is only based on one side. There is no verification of the base BTS after the MS has proved that it is permitted to connect to the network. This can lead to the MS being connected to a fake BTS because of the unfair reverse verification process by the BTS.

*D. Problems with GSM Network*

The mechanism known as A3/A8 is used to authenticate a mobile station on the GSM network. Actually, COMP128 is the key hash function used in the A3/A8 encryption algorithms.

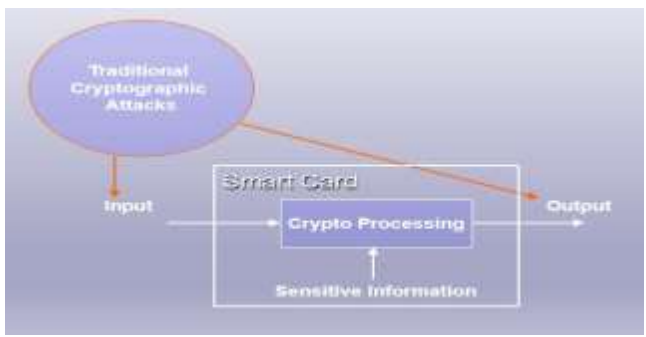

Fig. 3.4 Traditional Cryptographic Assumption
Source: [11]

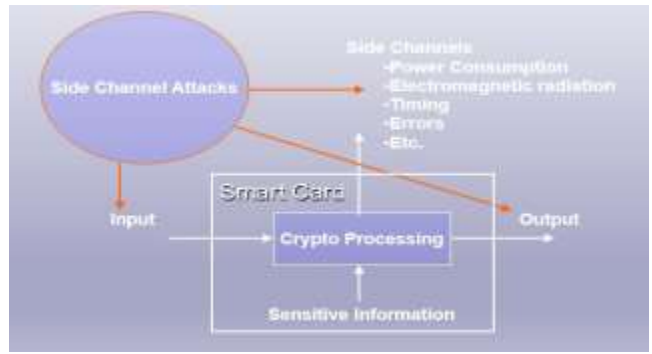

Fig. 3.5 Actual Information Available
Source: [11]

Partitioning attack was used on the COMP128. The goal of the attack was to find the value of Ki, knowing that if Ki is known, cloning the SIM card will be possible. The idea of the attack has to do with finding a violation of the cardinal principle. The violation of the cardinal principle means that signals on the side channels do not depend on output, input and sensitive information. Also, the statistical dependency in signals can be exploited to extract sensitive information (Max Stepanov, n.d).

In accordance with the closed GSM industry, the bug ridden A3, A5, and A8 are kept secret and undocumented. Eventually, Wagner identified a serious weakness, with Goldberg, when they proved that is was possible to obtain the Ki value with the use of reverse engineering, therefore making SIM cloning possible (Chikomo et al, n.d). According to Strobel 2007, Ian Goldberg and David Wagner released an article about SIM cloning ([ISA98]) in 1998. In the same vein, Alex Biryukov, David Wagner, and Adi Shamir in 2000 presented a white paper about real time cryptanalysis of A5/1 on a PC ([BSW01]). Though not in use as a standard anymore, the A5/2, successor of A5/1 is even more insecure.

*E. Attack on the A/5 Algorithm*

As stated above, the A5 encryption algorithm is used to provide mobile communication privacy by providing communication between mobile stations and the BTS. Kc is the Ki and RAND value fed into the A5 algorithm. The secret key (the Kc value) is used with the A5 algorithm for the encryption (Chikomo et al, n.d).

The A5/1, commonly used in the western hemisphere was deemed "strong encryption" till it was proven otherwise with a reverse engineering process. The A5/1's 64-bit keys are vulnerable to time memory tradeoff attacks, the key can be cracked with rainbow table. A normal computer with 2TB in table and 4 to 5 hard disks can crack the A5/1 key in 20 seconds or less (Karsten Nohl & Sylvain Manaut, 2010). Wagner and Goldberg also cracked the A5/2 key using a method that required five o'clock cycles to render A5/2 almost useless (Chikomo et al, n.d).

These successful attacks on A5 encryption algorithm are proofs that eavesdropping between mobile station and the BTS is still possible. This eavesdropping possibility now makes using GPRS over the GSM core network inappropriate and insecure especially for banks. (Chikomo et al, n.d).

*F. Security problem with SMS*

Initially, the idea behind the usage of SMS was to send an ordinary message over the GSM network. During the design of the GSM architecture, text encryption, mutual authentication, non-repudiation, and end-to-end security were omitted. This creates more problems than it solves. Since the phone directly processes SMS messages as soon as it gets it from the network, it opens a wide door for true remote bugs. Thus, the following are some of the problems inherent in SMS.

- *Forging Originating Address*
  This has to do with the altering of the originator's address field on the header of the SMS message to another alphanumeric string so as to hide the original senders address. This attack is known as SMS spoofing. It involves a third party sending out messages that appear to be from a genuine sender (Chimoko et al, n.d). It is can also be used with the help HLR query to determine a subscriber's true location anywhere on the globe.
- *SMS Encryption*
  SMS Security was not part of the GSM architecture design, hence the default data format for SMS is in plaintext. No end-to-end encryption is available. However, the only encryption available is the weak A5 encryption between the mobile station (MS) and the base transceiver station (BTS) (Chimoko et al, n.d).

One very serious weakness in the GSM network is the fact that lots of GSM traffic are predictable, therefore known key streams are provided (Nohl, 2010).

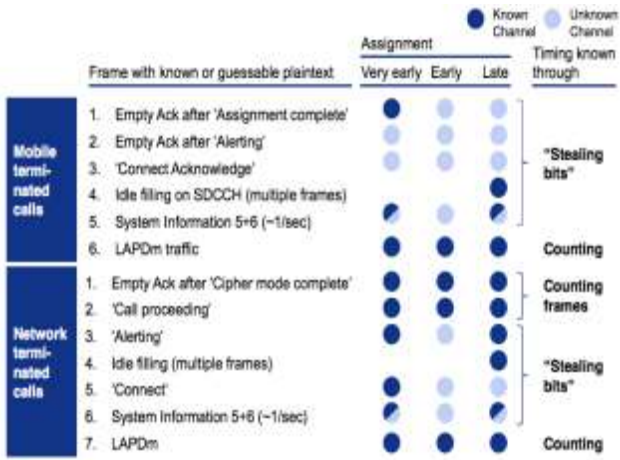

Source: [9]

## IV. RESULT AND DISCUSSION

*Security Requirements*

To fulfil the objective of this paper, which will addresses the security implications of the closed GSM system, interviews have been carried out on business pilots – Banks that use GSM system and other end users. Application scenarios have also been discussed to derive their security concerns and requirements.

Based on the scope of this paper, the security requirements will concentrate only on the mobile station and the air interface of the GSM network. The security requirements include among others: improved subscriber privacy and improved subscriber authentication. These will address the issue of connecting to a fake BTS or the control and retrieving of any information from the MS by the fake BTS. It will also address the issue of an unauthorized monitoring of the subscribers' traffic on the air interface (3G Partnership Project, 2000). From the user point of view, these security requirements are applicable to all digital air interfaces irrespective of the service providers.

*Improved Subscriber Authentication*

The fact that the network does not need to authenticate itself poses a serious risk of a subscriber connecting to a false BTS. Therefore, to address this, the improved subscriber authentication provides a 2-way authentication between the MS and the BTS by means of a cryptographically generated challenge-response for any subscriber's MS requesting connection to or service from the BTS. This will authenticate both the BTS and the MS, and then the authorized BTS can allow a connection to be established with a genuine MS. The improved subscriber authentication will through the control channels provide a mechanism to authenticate every message transmitted by the BTS or the MS, that may in one way or the other jeopardize the security of the subscriber. This authentication procedure minimizes the possible reuse of authentication signatures through a replay attack (3G Project, 2000).

This improved security measure will also provide a mechanism that will verify the authenticity of a MS through the data that represents a valid subscription of the MS. This mechanism will also authenticate the BTS.

*Improved Subscriber Privacy*

Implementing a better encryption algorithm over the air interface – wherein AES is a better candidate – protects the subscriber signalling and data against eavesdropping by attackers. On the other hand, if the GSM industry wants to still keep using the A3, A8, and A5 encryption algorithms for authentication, voice privacy and data privacy respectively, then certain conditions should be met. For instance, COMP128 hash functions should be changed to Message digest (SHA 2). The vulnerability inherent in COMP128 crypto hash function is what drives the power users of the GSM network to look elsewhere for security, which has led to a situation where organizations need some sort of multifactor and challenge response authentication system to ensure a multitier security of their transactions.

Secondly, there should be a mutual authentication between the MS and the BTS. Both ends should generate and authenticate each other with equal signed responses (SRES) before any connection can be established and communication begins. However, the GSM operators only authenticate a subscriber for billing purposes and not for subscriber privacy purposes.

## V. CONCLUSION

In conclusion, it was discovered in the literature that existing books and other publications focus on "user" or "system administrator" topics such as network deployment. There also exists scientific literature about the signal processing involved in GSM and optimizations of that. Other books explain the layer 3 protocol very well, but only from a theoretical point of view. Although GSM is has a public standard and maintained by organizations like ETSI, IrDA, etc. there are very few people besides the group of GSM baseband maker who actually understand the details of operation in a GSM mobile phone.

Analysing the situation in the GSM industry, it shows that digital Baseband ASICs and their corresponding software are present in billions of mobile phones, but the detailed knowledge on how they work is so far restricted to a small elite of engineers working for the industry. Compare that with the knowledge of the Internet protocols such as Ethernet, IP, TCP, HTTP, SMTP and others. Virtually every IT professional around the world understands them, and the knowledge is wide spread. One of the major reasons for that is the existence of no Free Software or Open Source software implementations.

## REFERENCES

[1]. Chamberlain, S. et al. (2010). The Red Hat newlib C Library. [Online PDF]. Available at: <ftp://sources.redhat.com/pub/newlib/libc.pdf> [Accessed 13 September 2016].
[2]. Gadaix, E. (2001). GSM and 3G security. [Online]. Available from www.blackhat.com/presentations/bh-asia-01/gadiax.ppt [Accessed 13 September 2016]
[3]. GNU (2011). GCC, the GNU Compiler Collection. [Online]. Available at: <http://gcc.gnu.org/> [Accessed 13 September 2016].
[4]. GNU (2011). GNU Binutils. [Online]. Available at: <http://www.gnu.org/s/binutils/> [Accessed 13 September 2016].
[5]. Lifchitz, R. (2010). Android Geolocation Using GSM Network. [Video Online] Available at: <http://www.youtube.com/watch?v=M-H2Ge2xUvg> [Accessed 13 September 2016].
[6]. Margrave, S. (n.d). GSM Security and Encryption [Online] Available at: <http://www.hackcanada.com/blackcrawl/cell/gsm/gsm-secur/gsm-secur.html> [Accessed 13 September 2016].
[7]. Microsoft. (2011). Safety and security centre. [Online]. Available from - https://www.microsoft.com/en-us/safety/online-privacy/public-wireless.aspx [Accessed 13 September 2016]
[8]. Mulliner C., & Golde, N. (2010). SMS-o-Death: From analysis to attacking mobile phones on a large scale. [Video Online]. Available at: <http://www.youtube.com/watch?v=8bkg3AjY6fs&feature=related> [Accessed 13 September 2016].
[9]. Nohl K. & Manaut S. (2010). Wideband GSM Sniffing. [PDF] Available at: <http://events.ccc.de/congress/2010/Fahrplan/attachments/1783_101228.27C3.GSM-Sniffing.Nohl_Munaut.pdf > [Accessed 13 September 2016].
[10]. Nohl, K. & Manaut, S. (2010). Wideband GSM Sniffing. [Online Video] Available at: <http://www.youtube.com/watch?v=lsIriAdbttc&feature=related> [Accessed 13 September 2016].
[11]. Osmocombb. (2009). OpenBSC. [Online]. Available from - http://openbsc.osmocom.org/trac/wiki/OpenBSC. [Accessed 13 September 2016]
[12]. Osmocon, (n.d.). Osmocon [Online]. Available at: <http://bb.osmocom.org/trac/wiki/osmocon> [Accessed 13 September 2016].
[13]. Rose India. (2008). WIFI Security for public networks. [Online]. Available at: <http://www.roseindia.net/technology/wifi/security-for-public-wifi.shtml>. [Accessed 13 September 2016]
[14]. Stipanov, M. (n.d). GSM Security Overview [PPT]. Available at: <www.cs.huji.ac.il/~sans/students_lectures/GSM%20Security.ppt > Accessed 13 September 2016]
[15]. TutorialsPoint, (n.d). GSM Architecture. [Online]. Available at: <http://www.tutorialspoint.com/gsm/gsm_network_switching_sub system.htm> [Accessed 13 September 2016].
[16]. Walte, H. and Markgrave, S. (2009). Project rationale. [Online]. Available from - http://bb.osmocom.org/trac/wiki/ProjectRationale. [Accessed 13 September 2016]
[17]. Weinmann, R., (2009). The Baseband Apocalypse: All your baseband are belong to us. [Video Online]. Available at: <http://www.youtube.com/watch?v=CPPQ8vA6cRc> [Accessed 13 September 2016].
[18]. Security Analysis Report. (2007). bridge-project. Retrieved 13 September 2016, from http://www.bridge-project.eu/data/File/BRIDGE%20WP04%20Security%20Analysis%20Report.pdf
[19]. *GSM*. (2015). *So.wikipedia.org*. Retrieved 13 September 2016, from https://so.wikipedia.org/wiki/GSM

*Bibliography*

[1]. Myagmar, G., 2001. 3G Security Principle. [Online]. Available from http://www.powershow.com/view/9f493-NzJhY/3G_Security_Principles_powerpoint_ppt_presentation [Accessed 13 September 2016]
[2]. Ortiz, A. & Prieto, A., n.d. SMS transimission using PDU mode and 7-bit coding scheme. [PDF Online]. Available at: <https://static.aminer.org/pdf/PDF/000/259/265/sms_transmission_using_pdu_mode_and_bit_coding_scheme.pdf > [Accessed 13 September 2016].
[3]. Riley, C. J., (2001). OpenBSC: Running your own GSM network. [Online]. Available from http://blog.c22.cc/2009/08/15/openbsc-running-your-own-gsm-network [Accessed 13 September 2016]
[4]. WAP WTAI (GSM), (1999). Wireless application protocol: Wireless Telephony Application Interface Specification. Wireless Application Forum: GSM Specific Addendum
[5]. Chikomo, K., Ki Chong, M., Arnab, A., & Hutchison, A. (2006). Security of Mobile Banking (1st ed.). Cape Town Rondebosch 7701, South Africa.